%% file: main.tex
\documentclass[journal,onecolumn]{IEEEtran}

\usepackage{amssymb,amsmath,paralist}
\usepackage{lipsum}
\usepackage[english]{babel}
\usepackage[utf8]{inputenc}
\usepackage{graphicx}
\usepackage[colorinlistoftodos]{todonotes}
\usepackage{amsfonts}
\usepackage{mathtools}
\usepackage{etoolbox}
\usepackage{mdframed}
\usepackage{enumitem}
\usepackage{algorithm}
\usepackage{algorithmicx}
\usepackage[noend]{algpseudocode}
\usepackage{multirow}
\usepackage{algorithm}
\usepackage{algorithmicx}
\usepackage[noend]{algpseudocode}
\usepackage{geometry}
\usepackage{float}
\usepackage{bbding}
\usepackage{pifont}
\usepackage{amssymb}

\DeclareMathOperator*{\argmax}{arg\,max}

\title{Thech. Report: Genuinization of Speech waveform PMF for speaker detection spoofing and countermeasures\footnote{This document was written in September 2023.}}

\author{
\IEEEauthorblockN{
Itshak Lapidot\IEEEauthorrefmark{1}\IEEEauthorrefmark{2},
Jean-Fran\c{c}ois Bonastre\IEEEauthorrefmark{2}\\}

\IEEEauthorblockA{\IEEEauthorrefmark{1}Afeka Tel-Aviv Academic College of Engineering, ACLP, Israel.}\\
\IEEEauthorblockA{\IEEEauthorrefmark{2}Avignon University, LIA, France.}\\

Email: \IEEEauthorrefmark{1}itshakl@afeka.ac.il,
\IEEEauthorrefmark{2}jean-francois.bonastre@univ-avignon.fr}

\allowdisplaybreaks

\begin{document}
\makeatletter
\maketitle

\begin{abstract}
\input abstract
\end{abstract}


\input introduction

\input genuinization_process

\input anti-spoofing_experiments

\input conclusions

\section*{Acknowledgment}
This work was supported by the ANR–JST CREST VoicePersonae project

\bibliographystyle{IEEEtran}
\bibliography{mybib}

\mbox{}

\makeatother
\end{document}

%% file: abstract.tex
In the context of spoofing attacks in speaker recognition systems, we observed that the waveform \textit{probability mass function} (PMF) of genuine speech differs significantly from the PMF of speech resulting from the attacks.
This is true for synthesized or converted speech as well as replayed speech. We also noticed that this observation seems to have a significant impact on spoofing detection performance. In this article, we propose an algorithm, denoted \textit{genuinization}, capable of reducing the waveform distribution gap between authentic speech and spoofing speech.

Our \textit{genuinization} algorithm is evaluated on ASVspoof 2019 challenge datasets, using the baseline system provided by the challenge organization. We first assess the influence of \textit{genuinization} on spoofing performance. Using \textit{genuinization} for the spoofing attacks degrades spoofing detection performance by up to a factor of $10$. 
Next, we integrate the \textit{genuinization} algorithm in the spoofing countermeasures and we observe a huge spoofing detection improvement in different cases. The results of our experiments show clearly that waveform distribution plays an important role and must be taken into account by anti-spoofing systems.

%% file: introduction.tex
\section{Introduction}
\label{sec:Introduction}

In recent years the sensitivity of speaker recognition to spoofing attacks and the development of spoofing countermeasures raised an increasing interest, \cite{Bonastre2007,Wu2012, Evans2014,Wu2015,Todisco2016,Wu2017,himawan2019deep}.
The most common threats in voice authentication are replaying recorded utterances, voice synthesis, and voice conversion. The associated countermeasures generally consist of a specific additional system capable of separating actual examples of speech and examples of impersonation, regardless of the type of impersonation attacks. Different approaches are applied, \cite{Todisco2017,Sriskandaraja2017,Sahidullah2015,Wu2016}. One of the main differences between these approaches (as well as between speaker recognition and spoofing detection) is related to the feature extraction. 
Different features were proposed for anti-spoofing systems \cite{Sahidullah2015}. Sometime, the feature used for spoofing detection is linked to the feature used for the speaker recognition task or is optimized together with it \cite{li2019anti,gomez2020joint}. As a result of the ASVSpoof challenge suite, the most promising appears to be \textit{constant Q cepstral coefficients} (CQCC) \cite{Todisco2017} which is a non-linear extension of the \textit{linear frequency Cepstral coefficients} (LFCC). Most of features offered are based on short-term spectral conversion (e.g., \textit{mel--frequency cepstral coefficients} (MFCC) and CQCC) and ignore the time domain. 
Moreover, even the rare exceptions that take the time domain into account usually only use it as a pre-processing step followed by short-term spectral analysis: \cite{Patel2017a} filters the voice excitation source in order to estimate the residual signal and uses it together with the frequency domain information inside a \textit{Gaussian mixture model} (GMM)-based classifier; \cite{Patel2017b} applies cochlear filtering and nerve spike density before to perform a short-term spectral analysis.

Spectral features are commonly used not only for countermeasures but also in many speech conversion and synthesis algorithm \cite{Wu2015a,Khodabakhsh2017}.

This apparent lack of interest in time domain information is surprising because time domain information is well known for its richness and is frequently used, for example, to estimate voice quality parameters and assess voice quality in clinical phonetics \cite{Verhelst1993,Deliyski1993,Alku1996,Gobl2003,Rusz2011,Gomez2009}. It seems obvious that at least the voice quality parameters are important for genuine or spoofing speech separation. If the time domain is mostly ignored in spoofing countermeasures, this is certainly more related to the intrinsic difficulty of time-based approaches than to the lack of information at this level.

In order to exploit temporal information without being drowned in the associated complexity, we have proposed in recent years to use a simple approach, the entropy of waveform coefficients. In \cite{Ben-Harush2009b} and \cite{Ben-Harush2009c} it has been shown that it allows to detect the overlap of speech between two speakers and in \cite{Lapidot2018a}, we successfully applied a similar approach to database assessment.

In  \cite{lapidot2019effects,lapidot2020effects}, we looked at an even simpler solution,  the \textit{probability mass functions} (PMFs) of waveform amplitude coefficients. We compared the PMFs of genuine voice recordings versus spoofing voice recordings (synthesized, converted or replayed voice) and were surprised by the large differences observed. We then exploited the waveform PMFs for the spoofing speech detection task.
We also proposed to correct the observed imbalance between authentic speech and speech spoofing PMFs, using a process inspired by the \textit{Gaussianization} of MFCC features \cite{Pelecanos2001}, applied at the waveform coefficient level and noted by analogy \textit{genuinization}.
In this article, we investigate further the effect of our \textit{genuinization} process when applied to different types of spoofing and genuine speech. We also question the behavior of \textit{genuinization} on high and low energized part of the speech signal and on spoofing detection system training set. 
Using this new knowledge on \textit{genuinization}, we propose a solution to make spoofing detection as much as possible insensible to the use of such a waveform manipulation.  All the experiments are done using the ASVspoof $2019$ challenge \cite{ASVspoof2019} train and the development sets as well as the baseline system provided by the challenge.

%% file: genuinization_process.tex
\section{Waveform PMF \textit{Genuinization} process}
\label{sec:Genuinization}

As noted in \cite{lapidot2019effects,lapidot2020effects}, the waveform coefficient PMFs can differ significantly between spoofing speech and genuine speech. This section is dedicated to a  transformation to be applied on the spoofing speech waveform coefficients to remove this difference. Our transformation is directly inspired from the data distribution conversion algorithm proposed in \cite{Pelecanos2001} for MFCC features and is denoted \textit{Genuinization}
as it targets the waveform PMF of genuine speech.

When in \cite{Pelecanos2001}, the targeted distribution is a Gaussian distribution (and the process is denoted \textit{Gaussianization}), we present here a generalized version of the algorithm to any continuous CDF then its adaptation to our \textit{Genuinization} case.

\subsection{Continuous CDF conversion}
\label {subsec:Data distribution_conversion}

\begin{figure} [ht!]
    \centerline{\includegraphics[width=\columnwidth]{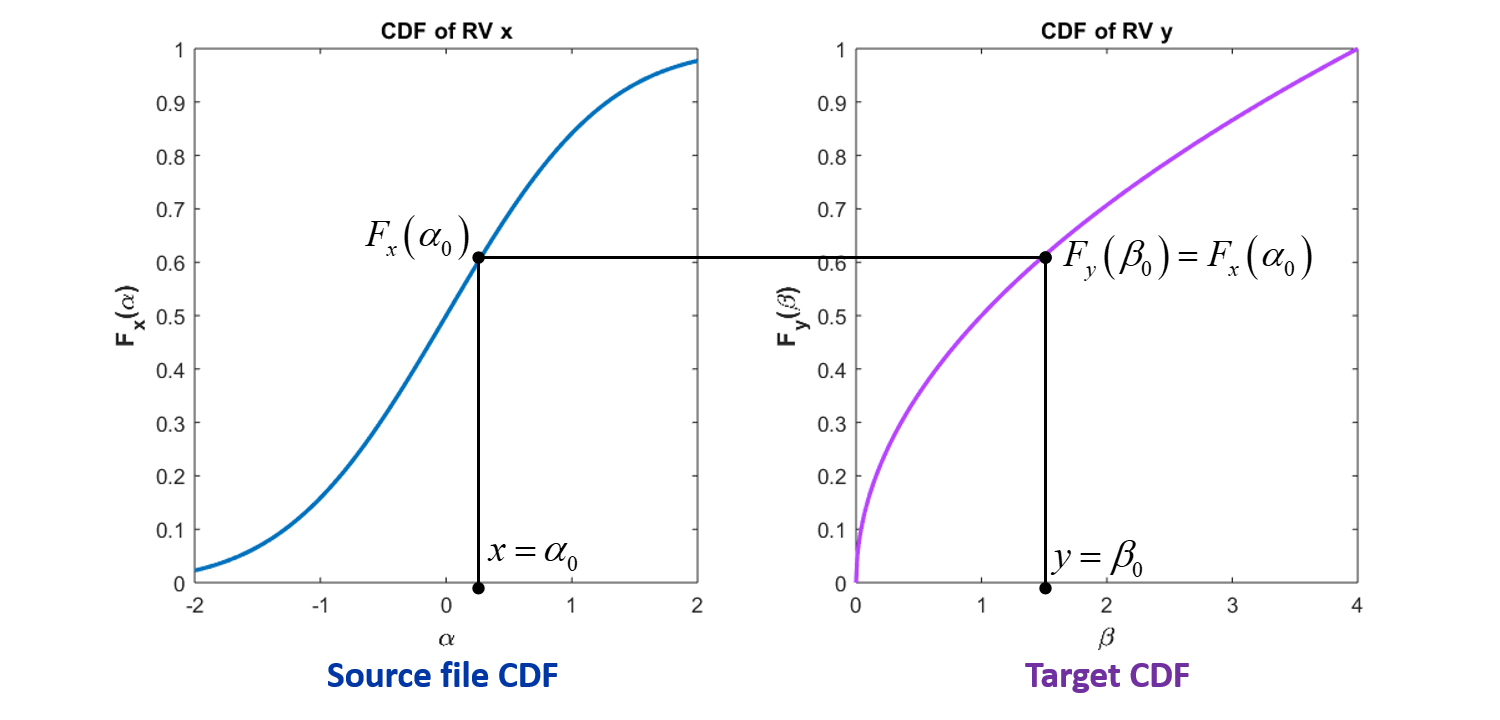}}
    \caption{\label{fig:DistributionConversion} Speech conversion illustration.}
\end{figure}

The main principle of the data distribution conversion algorithm is illustrated in Fig. \ref{fig:DistributionConversion}. Both source and target are continues \textit{random variables} (RV) denoted, respectively, $x$ and $y$. Their distributions are  represented by the corresponding \textit{cumulative distribution function} (CDF) denoted, respectively, $F_x$ and $F_y$. For $x = \alpha_0$, a data of the source, the algorithm:
\begin{itemize}
    \item Finds $F_x \left( \alpha_0 \right)$, the value of the source CDF for $\alpha_0$;
    \item Then, finds  $\beta_0$, the value of $F_y$, the target CDF, such as $F_y \left( \beta_0 \right) = F_x \left( \alpha_0 \right)$
    \item $y = \beta_0$ is the converted value of $x = \alpha_0$. 
\end{itemize}

\subsection{Basic \it{Genuinization} algorithm}
\label{subsec:Basic_Ginuinization_algorithm}
\label {subsec:Distribution_conversion}

The \textit{Genuinization} process aims to transform the samples of a source speech file in order to obtain a transformed speech file as close as possible to the original one but with a sample distribution following that of the genuine speech. The algorithm that we have just described is a simple and pleasant solution for this task, except that it is designed for the case of continuous random variables (in the \textit{Gaussianization}, the source is assumed to be continue and the target is a Gaussian distribution with zero mean and variance equal to 1). It is not valid for the  \textit{Genuinization} case, where the source and target variables are discrete\footnote{When it comes to discrete RVs, the use of the term CDF is not entirely appropriate and \textit{cumulative mass function} (CMF) is better suited. However, since CDF is more common to the reader, we will continue to use it for clarity in the rest of this article. Knowing that, we will assume that the CDF is not longer continue from the left. }: the data points are speech signals samples following a given quantization, on $16$bits here. It does not allow the one-to-one mapping used in the \textit{Gaussianization} case, even when the two distributions are similar.

\begin{figure} [ht!]
    \centerline{\includegraphics[width=\columnwidth]{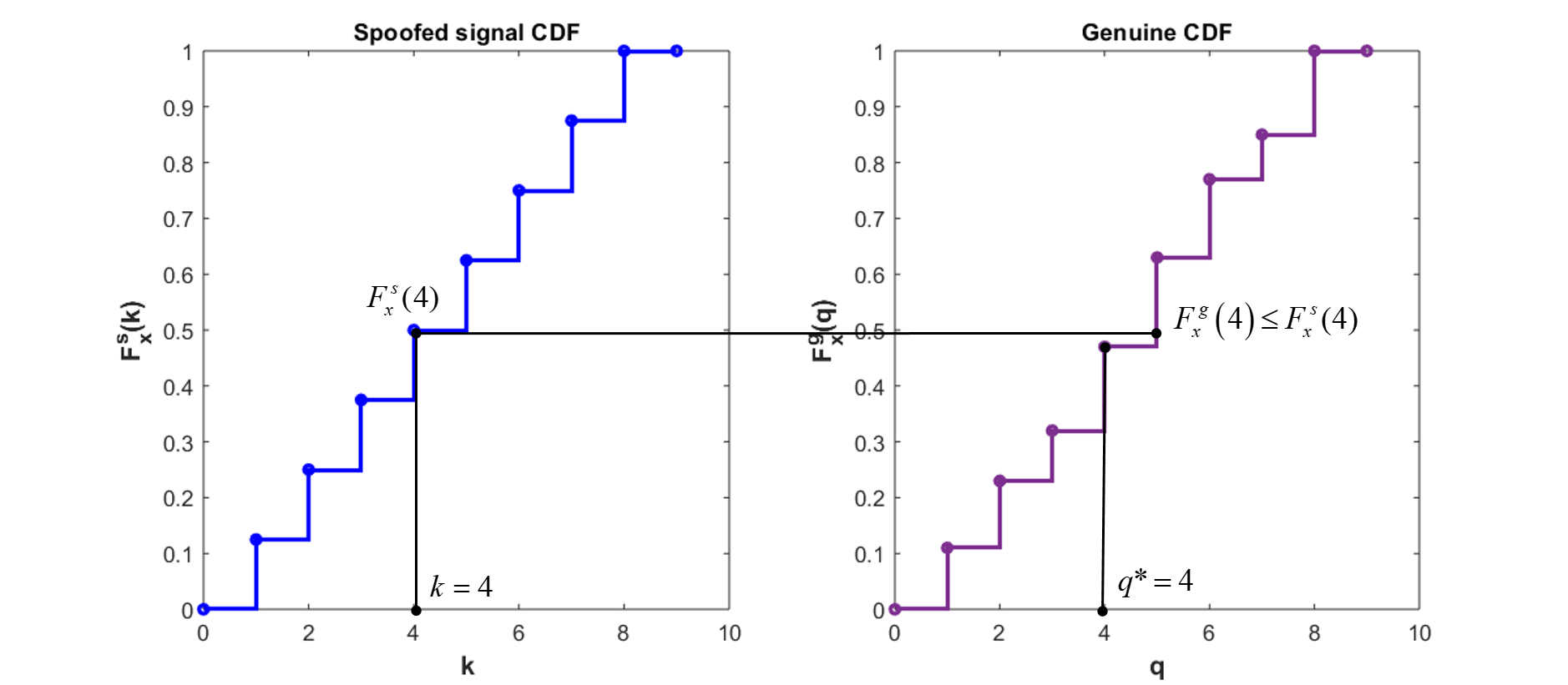}}
    \caption{\label{fig:Spoof2GenuineConversion_11} Speech conversion illustration for CDF of discrete RVs.}
\end{figure}

To overcome this limitation, we propose a quantile normalization \cite{hilger2002quantile} of the spoofing-speech sample distribution, using as target a genuine-speech sample distribution. Fig. \ref{fig:Spoof2GenuineConversion_11} and Alg. \ref{alg:Genuanization} illustrate the process. First, the genuine CDF, $F^{g}_x\left( k \right)$, is computed following $F^{g}_x\left(k\right)={\sum\nolimits_{q = 1}^{ k } {p^{g}_x\left( q \right)}}$, where ${p^{g}_x\left( k \right)}$ is the waveform PMF of the genuine speech.

where $s(n)$ is the spoofing file to \textit{genuinized}, 

Our \textit{Genuinization} is described bellow:
\begin{itemize}
    \item Given ,$x$, a discrete random variable $x \in \left\{1,\dots,2^{16}\right\}$, $g$, the genuine speech, $p^{g}_x\left(k\right)$ is its waveform amplitudes PMF.
    \item $k$ is the value that assigned to $x$ (the actual signal's amplitude is $s\left(n\right)=-1 + k \cdot 2^{-15}$).
\end{itemize}

Next, the CDF of genuine speech, $F^{g}_x\left(k\right)={\sum\nolimits_{q = 1}^{k} {p^{g}_x\left(q\right)}}$, is calculated over all the training set. For each spoofing speech signal $s\left(n\right)$, the CDF $F^{s}_x(k)$ ($s$ for the spoofed signal) is calculated in the same way as previously. The \textit{genuinization} algorithm is then applied, as described in Algorithm \ref{alg:Genuanization}, and illustrated in Fig. \ref{fig:Spoof2GenuineConversion_11}. For each data point of the spoofed signal $s(n)$, the value $k$ is found and $F_x^s \left( k \right)$ is assigned. Than, the genuine CDF value which is the closest from below to $F_x^s \left( k \right)$ is defined as an optimal match, $F_x^g \left( q\ast \right)$. $q\ast$ is assigned to be the genuinized value which will define the $\hat{s}\left(n\right)$. 

\begin{algorithm} [hbt!]
\caption{Genuanization algorithm}
\label{alg:Genuanization}
	\begin{algorithmic}
		\Require
        \Statex Given a spoofing file, $s\left(n\right)$
 		\Comment{$n=1,\dots,N$}
        \Statex Be the genuinized file, $\hat{s}\left(n\right)$
        \Statex Genuine CDF $F^{g}_x\left(k\right)$
 		\Comment{$k\in 1,\dots, 2^{16}$}
 		\Statex Spoofing file CDF $F^{s}_x\left(k\right)$
		\For{$n:=1$ \textbf{to} $N$ \textbf{step} $1$}
            \State Set $k=\left[s\left(n\right)+1\right]2^{15}$.
            \State Find $q^{\ast} = \mathop {\argmax }\limits_q \left\{ {F_x^g\left( q \right) \leq F_x^s\left( k \right)} \right\}$
            \State Set $\hat{s}\left(n\right) = -1+2^{-15} \cdot q^{\ast} $
		\EndFor
	\Statex \textbf{Return:} $\hat{s}\left(n\right)$
\end{algorithmic}
\end{algorithm}

\begin{figure} [ht!]
    \centerline{\includegraphics[width=\columnwidth]{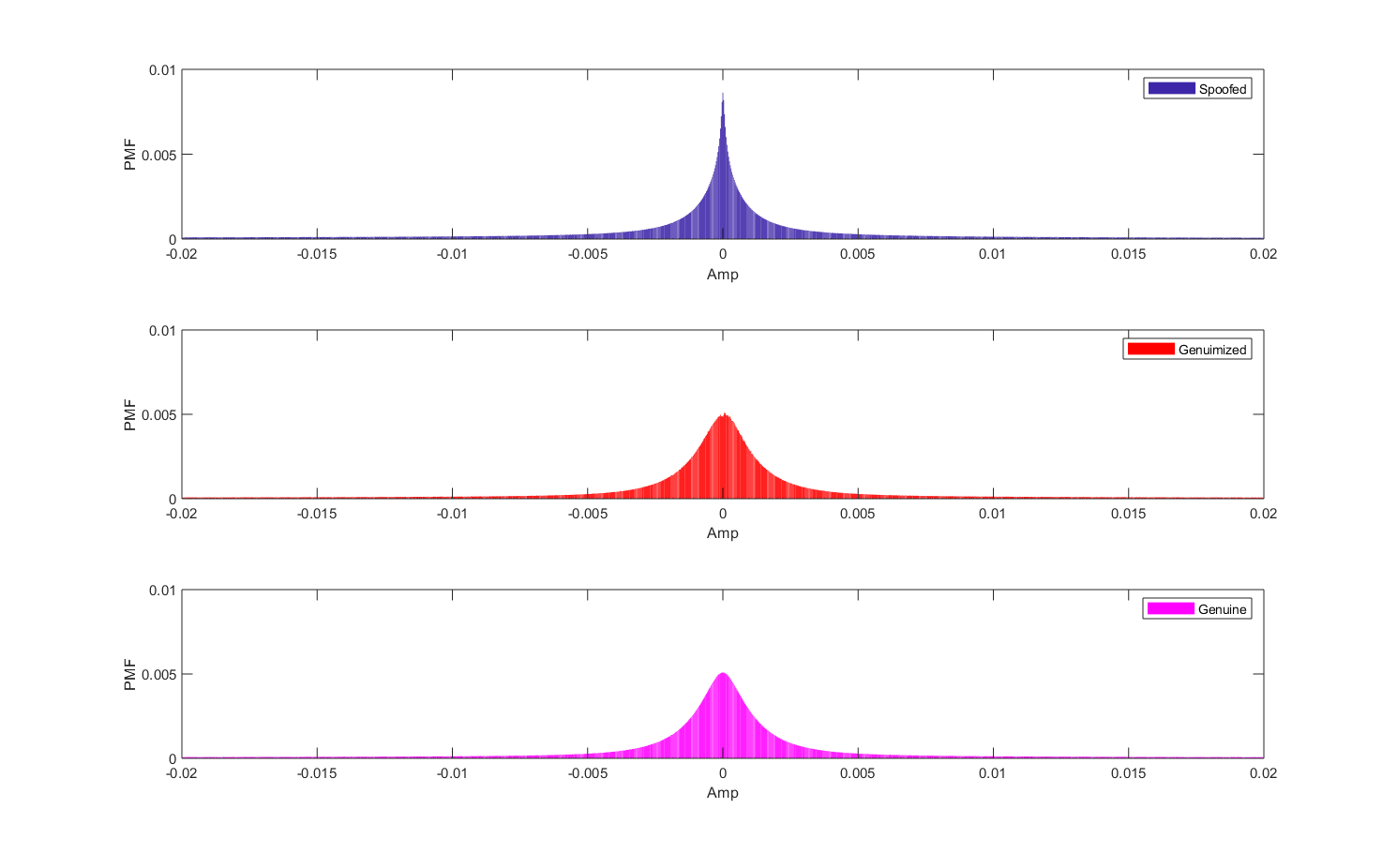}}
    \caption{\label{fig:trainDataPMFs_LA} Waveform amplitude PMFs for logical condition, train set: Spoofed (upper); Genuinized (middle); Genuine (bottom).}
\end{figure}

\begin{figure} [ht!]
    \centerline{\includegraphics[width=\columnwidth]{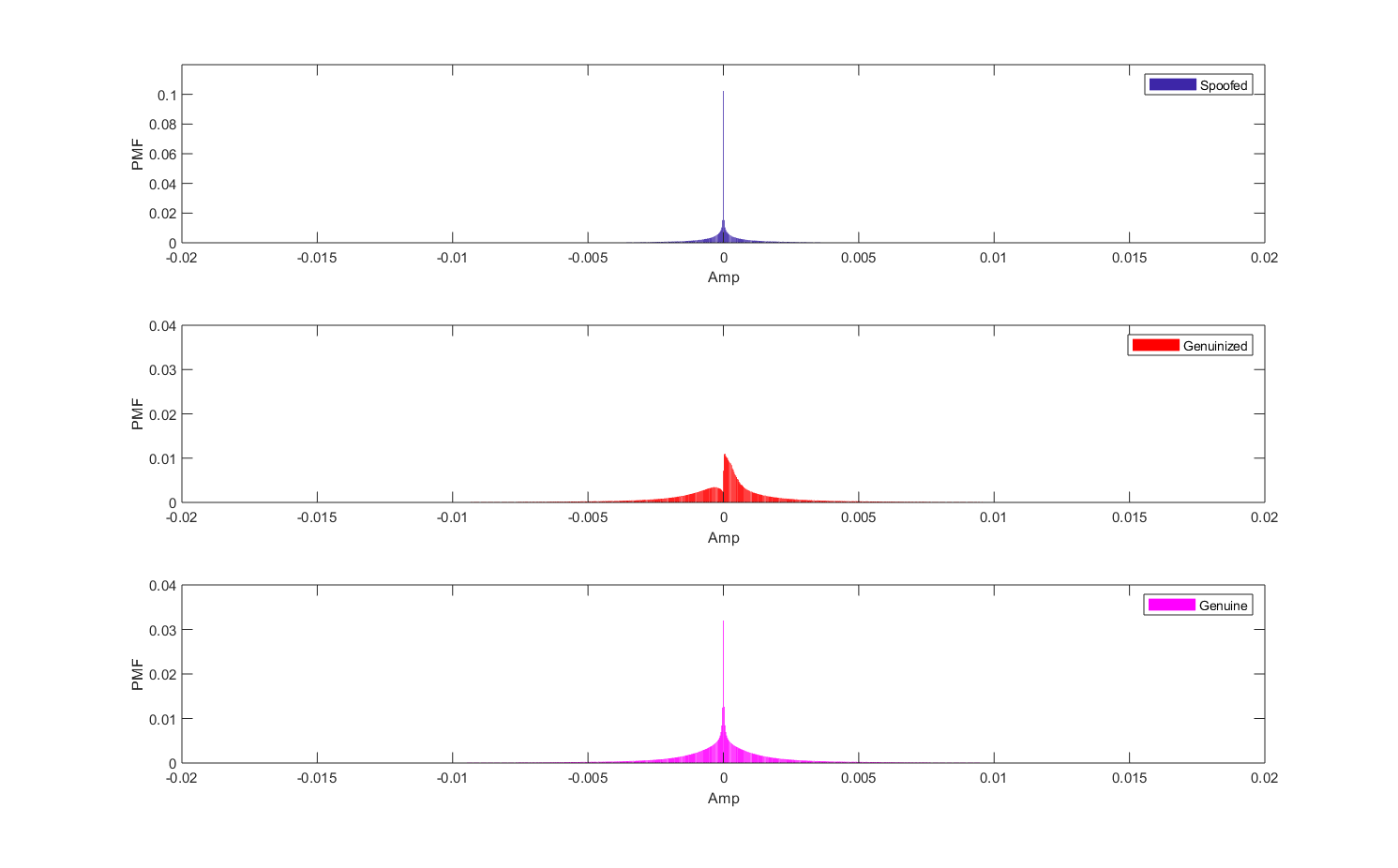}}
    \caption{\label{fig:trainDataPMFs_PA} Waveform amplitude PMFs for PA condition, train set: Spoofed (upper); Genuinized (middle); Genuine (bottom).}
\end{figure}

\begin{figure} [ht!]
    \centerline{\includegraphics[width=\columnwidth]{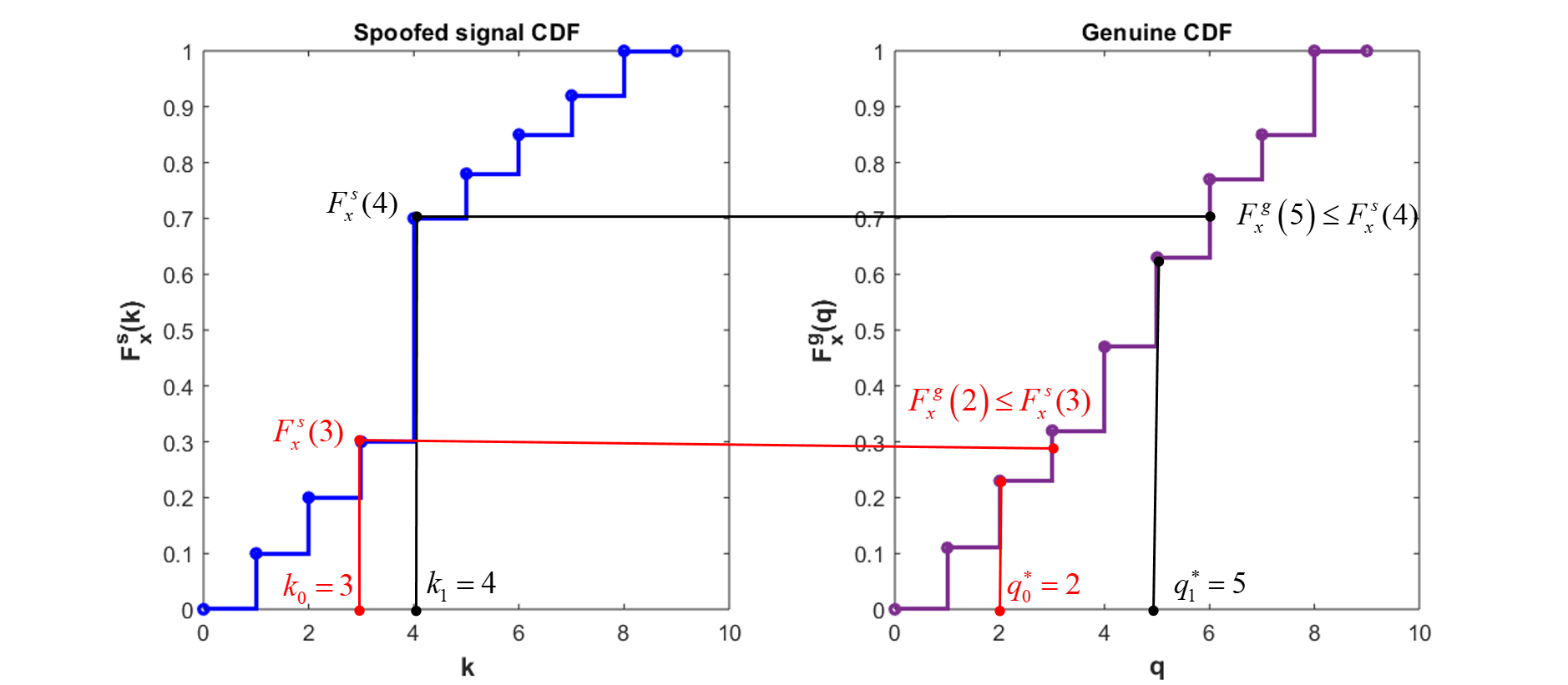}}
    \caption{\label{fig:Spoof2GenuineConversion_21} Speech conversion illustration for CDF of discrete RVs with high pick in the source PMF (spoofed signal PMF).}
\end{figure}

We applied this approach to genuinization of two databases of the ASVspoof2019 challenge~\cite{ASVspoof2019}. At this challenge there were two separated sub-challenges. The first is \textit{logical access} (LA) and it includes synthesis speech and converted speech; the second was \textit{physical access} (PA) and it includes replayed speech (this database is a simulated replayed database in order to have controlled conditions, both for recorded and replayed signals). In each challenge there was a train set and development set to design systems which classify for genuine or spoofed speech. The genuinization was applied on the test sets of both datasets. The genuine PMF and CDF were estimated from all genuine files at each dataset, while the genuinization was performed file by file and the overall PMFs of the spoofed files before and after genuinization were calculated. First, the genuinization was applied on LA dataset and the results are presented in Fig. \ref{fig:trainDataPMFs_LA}. The horizontal axis is the amplitude of the speech signal and not $k$. The upper sub-plot is the PMF of the spoofed data. At the middle it is the spoofed data after genuinization, while at the bottom is the PMF of the genuine speech which was used as a target distribution. It is clear that while the spoofed PMF significantly differs from the genuine PMF, the genuinized PMF is similar to the target, genuine PMF. Next, the same procedure is applied to the PA dataset and the results are presented in Fig. \ref{fig:trainDataPMFs_PA}. As the peek of the spoofed PMF in the PA case is much higher than the other two plots, the vertical axis have different scales ($0.12$ for spoofed PMF and $0.04$ for the others). In this case the genuinization algorithm dose not work, and a notch is observed at amplitude equals zero after genuinization. The main difference to the LA case is the spoofed PMF has a very high pick at one amplitude. Such case illustrated in Fig. \ref{fig:Spoof2GenuineConversion_21}. In the illustration the pick appears at $k=4$. Due to this peak, $q=3$ and $q=4$ never present in the genuinized distribution despite there presence in the target distribution. Such phenomena can appear only in discrete distributions as they CMF is not continuous from the left at the discrete values which have the probability mass. For this reason, the algorithm has to be adapted to perform also in such cases.

\subsection{Perturbated Genuinization algorithm}
\label{subsec:Randomized_Ginuinization_algorithm}

\begin{figure} [ht!]
    \centerline{\includegraphics[width=\columnwidth]{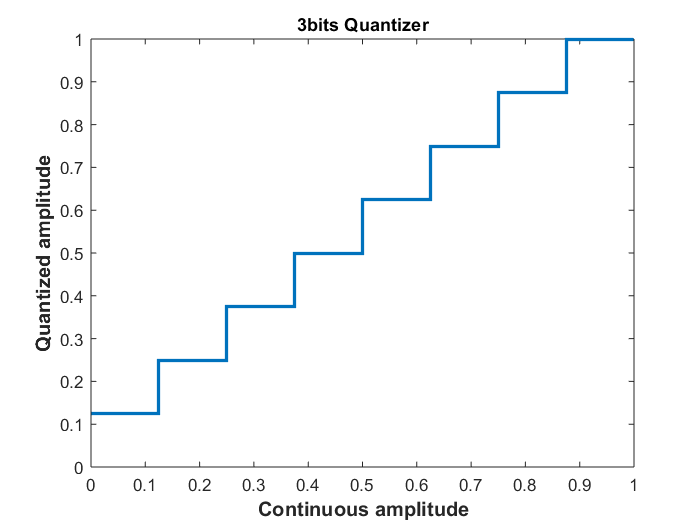}}
    \caption{\label{fig:3bitsQuantaizer} $3$bits quantizer.}
\end{figure}

\begin{figure} [ht!]
    \centerline{\includegraphics[width=\columnwidth]{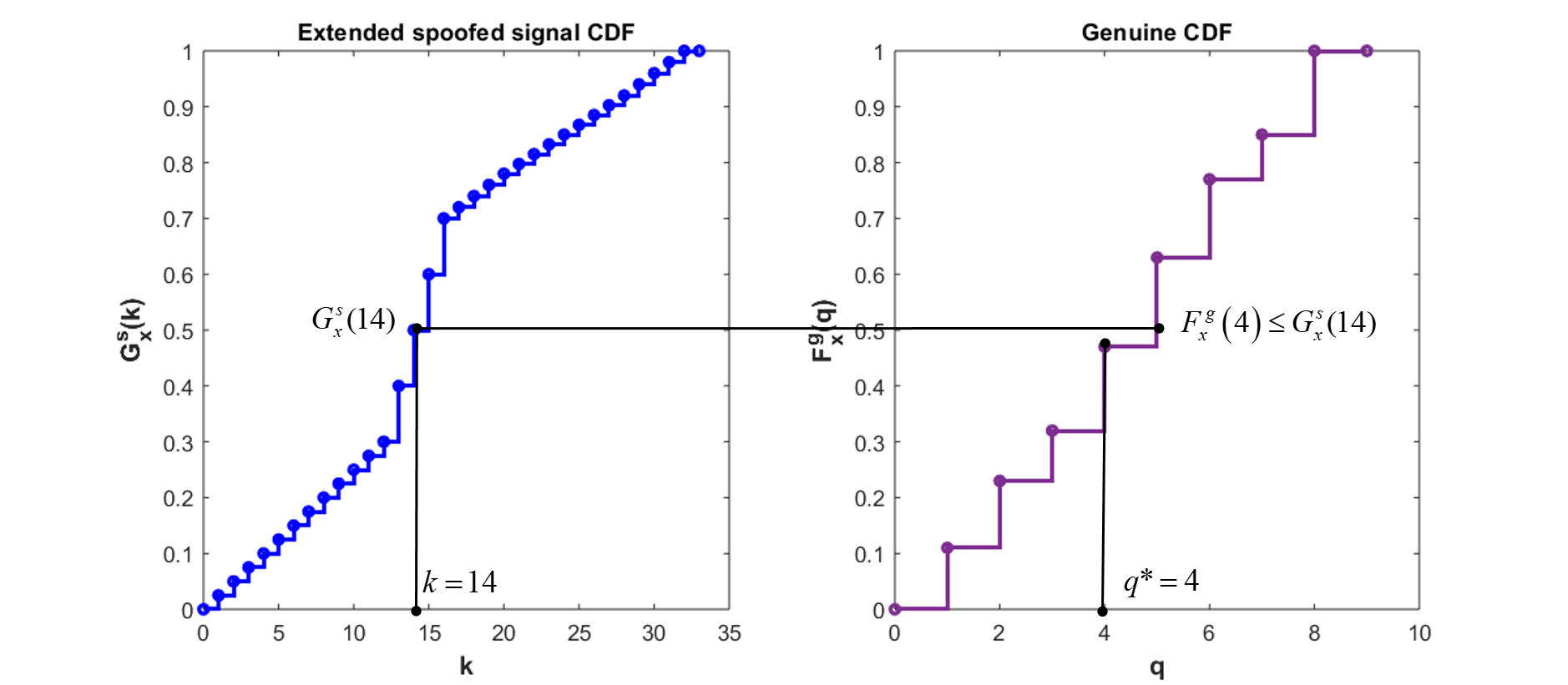}}
    \caption{\label{fig:Spoof2GenuineConversion_31} Speech conversion illustration for extended CDF of discrete RVs with high pick in the source PMF (spoofed signal PMF).}
\end{figure}

Before explaining the algorithm, it is important to remind the quantization function during the sampling process. In most databases the quantization is $16$bit per sample in the range $\left[ -1.0,1.0 \right]$, when the lowest level is $-1 + 2^{-15}$ and the highest level is $1.0$. In general case, for $D$-bits case, the quantization levels are between $-1+2^{-\left( D - 1 \right)}$ and $1.0$. In Fig. \ref{fig:3bitsQuantaizer} a $3$bits quantizer is presented. It is a mapping of many to one, when a segment $S_k = \left( -1+ \left( k - 1 \right) \cdot 2^{-\left( D - 1 \right)}, -1+  k \cdot 2^{-\left( D - 1 \right)} \right]$ mapped to its maximum value $-1+  k \cdot 2^{-\left( D - 1 \right)}$. This means that the probability of the quantization level at $k=K$, $p_K= \int_{{S_k}} {{f_A }\left( \alpha  \right)d\alpha }$, when ${f_A }\left( \alpha  \right)$ is the \textit{probability density function} (\textit{pdf}) of the continuous amplitude $A$. As the true $f_A \left( A \right)$ is not known and assuming that $S_k$ is sufficiently small, it is possible to approximate this \textit{pdf} as a piece-wise constant, $\hat{f}_A \left( \alpha \right) = \frac{p_k}{\Delta}$ where $\Delta = \frac{2}{2^D} = 2^{- \left( D - 1 \right)}$ (segments length), and $A \in S_k$ ($k$-th segment). According to these assumptions it is possible to find a new PMF and the CDF for quantizer with $\left( D + d \right)$-bits. Adding $d$-bits, means, adding $2^d - 1$ quantization levels to each segment (in total $2^d$ quantization levels per segment). The extended CDF is calculated only to the spoofed signal CDF, i.e., from $F_x^s \left( k \right)$ we obtained an extended spoofed CDF $G_x^s \left( k \right)$. Algorithm \ref{alg:Genuanization} can now be applied but with the new CDF, as shown in Fig. \ref{fig:Spoof2GenuineConversion_31}.

\begin{figure} [ht!]
    \centerline{\includegraphics[width=\columnwidth]{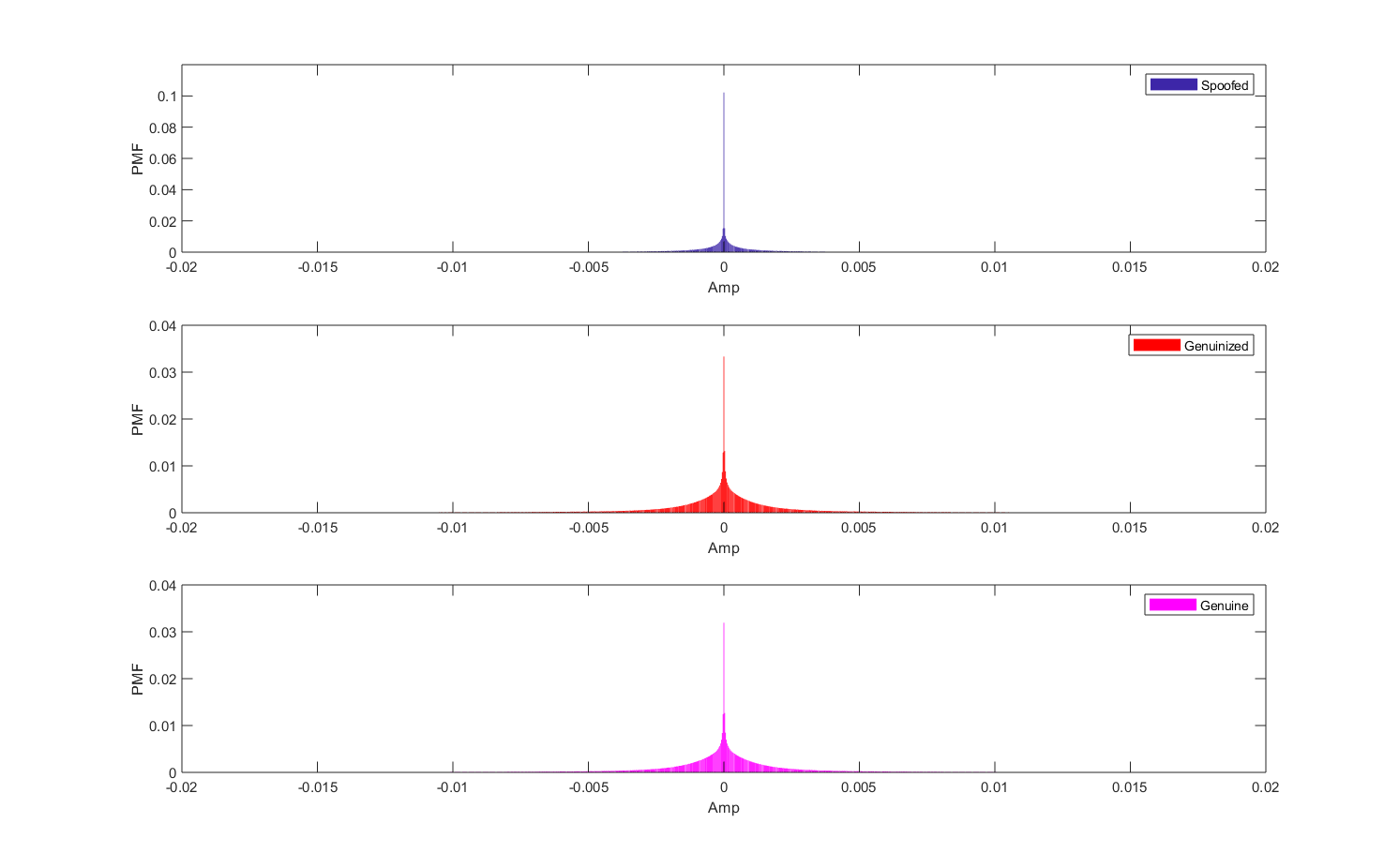}}
    \caption{\label{fig:trainDataPMFs_PA_Fixed} Waveform amplitude PMFs for PA condition, train set, using the randomized genuinization: Spoofed (upper); Genuinized (middle); Genuine (bottom).}
\end{figure}

The only unsolved problem is how to assign a new $k$ for $G_x^s \left( k \right)$. As $k$ represent a value of a signal in the $k$-th segment, and the assumption is of uniform distribution in a segment, the new $k$ is assign by uniformly distributed discrete random value between the $2^d$ new levels: $\left( k \cdot 2^d - n \right) \to k$, where $n$ is a discrete uniform random variable, $n \sim U \left( 0, 2^d-1 \right)$. As was mentioned above, in all the databases $D=16$ and we augmented it with $d=5$, extended the quntizer by a factor of $32$. The results after such manipulation is presented in Fig. \ref{fig:trainDataPMFs_PA_Fixed}. Comparing this result with the one in that presented in Fig. \ref{fig:trainDataPMFs_PA} shows a significant improvement. The results for LA remained the same. This algorithm is much more robust than the original genuinization algorithm, but the hyper-parameter $d$, the additional number of bits, have to be chosen in advance. Too small $d$ and the results will be insufficient, while too large $d$ will lead to overload in memory and the analysis time without loss in the end performance.

\subsection{The importance of the non-speech parts}
\label{subsec:Non-apeechParts}

\begin{figure} [ht!]
    \centerline{\includegraphics[width=\columnwidth]{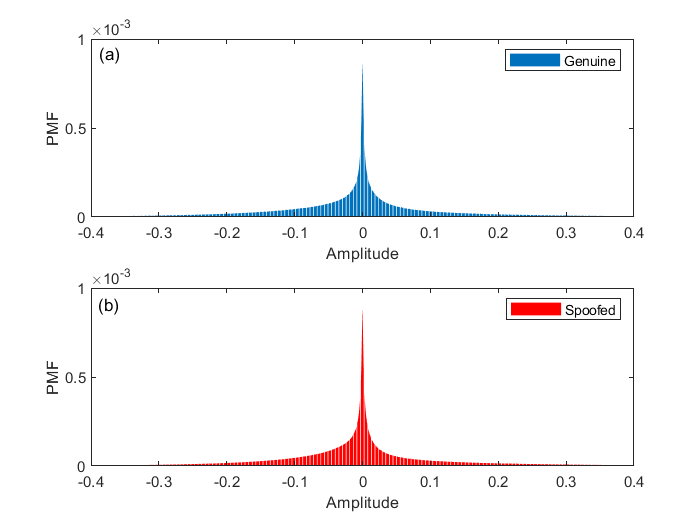}}
    \caption{\label{fig:LogicalAmpPMFwithVAD} Waveform amplitude PMFs for logical condition after VAD, train set: genuine speech (upper) and spoofed speech (bottom).}
\end{figure}

Another issue, is the question of what is learned by the anti-spoofing systems. Current spoofing countermeasure systems typically use the entire signal, including the non-voice parts~\cite{Valenti2018}. When a voice activity detector (VAD) is used to remove these unvoiced portions of the signal, performance is degraded. This result suggests that countermeasure systems pay great attention to parts of the signal where there is no speech-related information. If this is true, countermeasure systems appear sensitive to a change in PMF in the low amplitude region, that is, at the low energy portions of the signal and primarily at the non-speech portions.
In order to assess this sensitivity of spoofing detection systems to the low energy signal parts, we are conducing several experiments using VAD. We apply here the simple energy based VAD used in~\cite{Lapidot2018a,Ben-Harush2012} (using $\alpha  = 0.03$). The PMFs of the genuine speech and spoofed speech after removing the non-speech parts thanks to the VAD are shown in Figure~\ref{fig:LogicalAmpPMFwithVAD}. The PMFs looks very similar one to the other. 

In order to better understand whether our  \textit{genuinization} process is sensitive to the speech/non-speech question, we propose an experiment where the \textit{genuinization} parameters are learned using only non-speech parts. Other than that, there is no difference in the the procedure described in Algorithm~\ref{alg:Genuanization} (in other words, in the genuinization process, we change only the target distribution, which is now computed only on the non-speech segments detected by the VAD). 

\begin{figure} [ht!]
    \centerline{\includegraphics[width=\columnwidth]{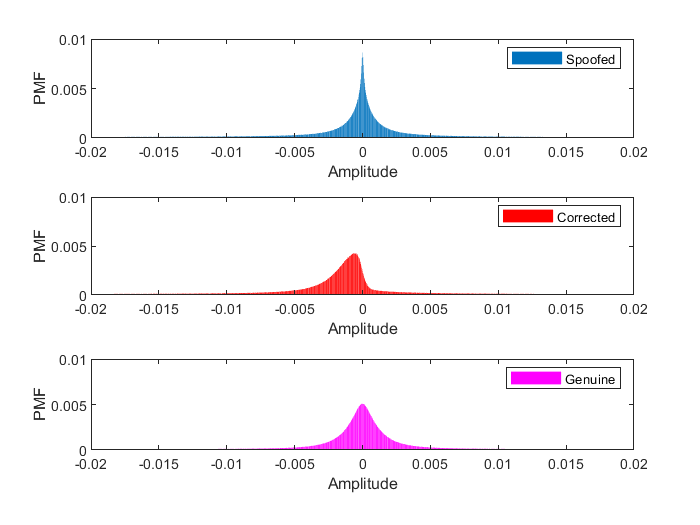}}
    \caption{\label{fig:SpoofedFixedPMF_nonSpeech} Waveform amplitude PMFs for logical condition, train set: original spoofing speech (upper), spoofing speech after non-speech-based genuinized (middle) and genuine speech (bottom).}
\end{figure}

\begin{figure} [ht!]
    \centerline{\includegraphics[width=\columnwidth]{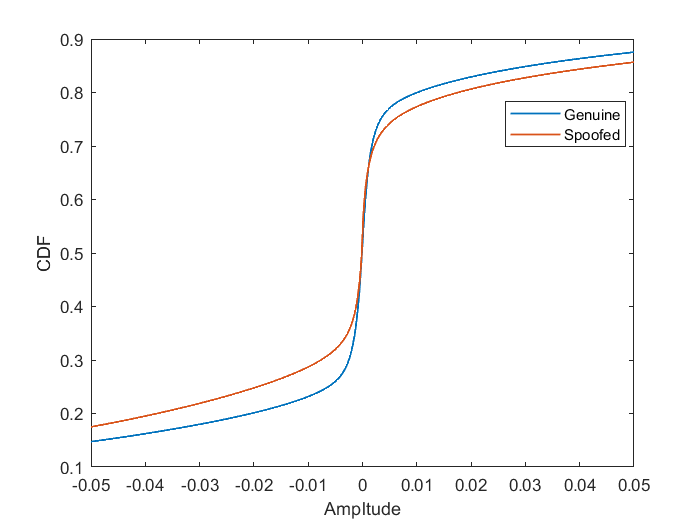}}
    \caption{\label{fig:GenuineSpoofedCDFs} CDFs of the Logical conditions train set waveform amplitudes for the Genuine and Spoofing speech.}
\end{figure}

 Figure~\ref{fig:SpoofedFixedPMF_nonSpeech} presents the PMFs of genuine speech, original spoofing speech and spoofing speech after this ''non speech-only'' \textit{genuinization}. The PMF after \textit{genuinization} is far from being identical to the genuine speech PMF. It is very clear when Figure~\ref{fig:SpoofedFixedPMF_nonSpeech} is compared with Figure~\ref{fig:trainDataPMFs_LA}, where the PMF after \textit{genuinization} is closer to the genuine speech PMF. This is explained by the heavier tails of spoofing speech, particularly the left tail, as showed in Figure~\ref{fig:GenuineSpoofedCDFs}. 

 Figure~\ref{fig:SpoofedFixedPMF_nonSpeech} presents the PMFs of genuine speech, original spoofing speech and spoofing speech after this ''non speech-only'' \textit{genuinization}. The PMF after \textit{genuinization} is far from being identical to the genuine speech PMF. It is very clear when Figure~\ref{fig:SpoofedFixedPMF_nonSpeech} is compared with Figure~\ref{fig:trainDataPMFs_LA}, where the PMF after \textit{genuinization} is closer to the genuine speech PMF. This is explained by the heavier tails of spoofing speech, particularly the left tail, as showed in Figure~\ref{fig:GenuineSpoofedCDFs}. 

\subsection{Random Genuinization}
\label{subsec:RandomGenuinization}

In random genuinization, the procedure is exactly the same as presented in section \ref{subsec:Randomized_Ginuinization_algorithm}, with one exception: the reference PMF is not estimated from an entire training set, but each time a random file is selected. In such a manner, not all the genuinized recordings will have almost the same PMF. It is a more realistic distribution, when each recording is having its own PMF and not using a predefined collective PMF.

%% file: anti-spoofing_experiments.tex
\section{Impact of genuinization on the spoofing countermeasure}
\label{sec:GenuinizationImpact}

In this section we present the effects of the genuinization and random genuinization on the countermeasure. We apply the genuinization on the development set of ASVspoof, 20219~\cite{ASVspoof2019}. As in such application there are always two sides, the attaker and the countermeasure. On the attacker side, the PMF model for genuinization was estimated from the genuine speech of the evaluation set of ASVspoof~2019. It is important to emphasize that on the attacker side the genuinization was performed only on spoofed speech, while genuine speech stayed untouched. On the countermeasure side, there are two actions that can be taken:
\begin{enumerate}
    \item Training the GMMs for the genuine spoof model using genuinized speech. In this case, the genuine speech PMF estimation is done from the genuine speech of the ASVspoof~2019 train set. All the spoofed speech recordings of the train set are genuinized, and served for training the spoofed GMM. When genuine speech genuinization is performed to train the genuine GMM, the same PMF was used as for spoofed speech, i.e., same data was used both for PMF estimation and for genuinization.
    \item On the countermeasure side it is possible to genuinized all the files under examination (the development set), but it must be done for all the recordings, as the countermeasure does not have the information whether the recording is genuine speech or spoofed neither the information whether the spoofed recordings where genuinized or not. The genuinization performed using the PMF estimated on the train, genuine speech data, exactly the same used for GMMs training.
\end{enumerate}
When we describe the actions on both, attacker and countermeasure sides, the genuinization can be as described in subsection~\ref{subsec:Randomized_Ginuinization_algorithm} or random, genunianization in section genuinization, subsection~\ref{subsec:RandomGenuinization}.

The following experiments test the preferred action that countermeasure can take in the situation of lack of knowledge whether the genuinization was performed by the attacker or not, and which genuinization, random or not. The condition notations for the attack and the countermeasure defined in table~\ref{table:Notation}. Each recording can belong to human$\backslash$genuine speech or spoofed speech and belongs to a train set or to test set; the GMM of a spoofed speech can be trained on spoofed data without genuinization, after genuinization, or after random genuinization; attacker action is applied only to spoof recordings and can be no action, genuinization or random genuinization. While the countermeasure action is applied to all the files under examination. It means that \textbf{AC} may have $9$ different options: NN, NG, NR, GN, GG, GR, RN, RG, and RR, for both, train and test.

\begin{table}[ht]
\centering
\caption{\label{table:Notation}{Notation for countermeasure experiments}}
    \begin{tabular}{cc}
        \hline
        \textbf{H} & Human$\backslash$Genuine\\
        \textbf{S}  & Spoof\\
        \textbf{Tr} & Train\\
        \textbf{Te} & Test\\
        \hline
        \multicolumn{2}{c}{\textbf{Countermeasure}}\\
        \textbf{O} & Original GMM\\
        \textbf{G} & Genuinezed GMM\\
        \textbf{R} & Randomly Genuinized GMM\\
        \hline
        \multicolumn{2}{c}{\textbf{AC - Attacker action Countermeasure action}}\\
        \textbf{N} & No action\\
        \textbf{G} & Genuinized recording\\
        \textbf{R} & Randomly genuinized recording\\
        \hline
    \end{tabular}
\end{table}

For the train conditions the training of the genuine GMM, $\bf{H\_Tr}$ and the Spoofed GMM, $\bf{S\_Tr}$ can be either \textbf{O}, \textbf{G}, or \textbf{R}. It ends for $9$ different possibilities. However, $4$ possibilities are illogical and were not examined: $\bf{H\_Tr} = G or R$, while $\bf{S\_Tr = O}$. There is no reason the use a genuine speech model after genuinization, if the spoof model is the original one. Also $\bf{H\_Tr} = G \setminus R$, while $\bf{S\_Tr = R \setminus G}$, i.e., different genuinization methods. At the end, there are only $5$ possibilities. On the side of the test ($\bf{H\_Te}$ and $\bf{S\_Te}$), the action that is taken by the attacker is always on the spoofed data, while the genuine speech is untouched, i.e., \textbf{N}. It leads to $3$ possibilities. From the countermeasure side, the action must be applied to all the data, as it is not known whether it is genuine or spoofed. It leads also to $3$ possibilities. This ended with $3 \times 3 = 9$ possibilities at the countermeasure side. In total, there are $5 \times 9 = 45$ possible scenarios which have to be evaluated for LA conditions and $45$ for PA conditions. Each experiment performed for LFCC and CQCC features.

As far as we know, it is a pioneer research in time domain. As such, the importance of this work is not to show that is makes the detection of spoofed speech harder or easier, but to show that it changes the behavior of the systems. In table~\ref{table:LA_PA_GenuineGenuinized} the results of both LA and PA conditions are presented for for original GMMs for both, genuine and spoofed speech.

\begin{table}[!ht]
  \centering
   \caption{\label{table:LA_PA_GenuineGenuinized}{LA and PA EER [\%]: The GMMs are trained on the original data, while the test data examined for all described manipulations.}}
   
  \begin{tabular}{r|cccc|cc||cc}
    & & & & & \multicolumn{2}{c}{\textbf{\textit{LA}}} & \multicolumn{2}{c}{\textbf{\textit{PA}}}\\
  	\hline
  	& $\bf{H\_Tr}$ & $\bf{S\_Tr}$ & $\bf{H\_Te}$ & $\bf{S\_Te}$ & \textbf{LFCC} & \textbf{CQCC} & \textbf{LFCC} & \textbf{CQCC}\\
	\hline
	$\bf{1}$ & O & O & NN & NN & $2.710$ & $0.430$ & $11.960$ & $9.870$\\
	$\bf{2}$ & O & O & NN & GN & $1.570$ & $3.336$ & $18.887$ & $11.924$\\
	$\bf{3}$ & O & O & NN & RN & $1.720$ & $4.749$ & $20.797$ & $11.630$\\
	$\bf{4}$ & O & O & NG & NG & $12.090$ & $5.850$ & $14.480$ & $20.370$\\
	$\bf{5}$ & O & O & NG & GG & $12.671$ & $6.083$ & $14.333$ & $20.479$\\
	$\bf{6}$ & O & O & NG & RG & $12.833$ & $6.476$ & $14.669$ & $20.575$\\
	$\bf{7}$ & O & O & NR & NR & $14.200$ & $8.630$ & $15.760$ & $22.500$\\
	$\bf{8}$ & O & O & NR & GR & $14.560$ & $8.163$ & $15.403$ & $22.519$\\
	$\bf{9}$ & O & O & NR & RR & $14.560$ & $8.791$ & $15.333$ & $22.647$\\
	\hline
  \end{tabular}
\end{table}

From table ~\ref{table:LA_PA_GenuineGenuinized} can be learned that when the countermeasure takes the action of genuinization (either regular or random) the results are practically independent from the action of the attacker (experiments $4-6$ and $7-9$) both for LA and PA conditions. Those observations are valid for all other combinations of GMMs. The values of the EER changes, but not the tendency. Another observation is that countermeasure action of of genuinization is at least as good as random genuinization in all the scenarios. Probably due to the larger statistic for PMF estimation, which makes it more robust to a larger variety of manipulations.

When applying the original data GMMs and do not take action on the countermeasure side, the results are usually the best if the attacker do not take any action as well ($1st$ experiment for both LA and PA). However, there is a significant degradation in EER if the attacker perform any kind of genuinization. The only exception is for LA conditions with LFCC features (experiments $2$ and $3$). In general experiments $4-9$ gave relatively poor results. It leads to a conclusions that the best not to make any manipulation on the data on the countermeasure side.

Next experiment repeats experiment $1-3$ from table~\ref{table:LA_PA_GenuineGenuinized}, but with different trained GMMs for both, genuine and spoofed speech. The results are presented at table \ref{table:GMM_GenuineGenuinized}. From examining experiments $4-12$ and comparing the experiments $1-3$, which are the same as in table~\ref{table:LA_PA_GenuineGenuinized}, it can be observe that testing with original GMMs when no action is taken by both side is significantly better than using at least one GMM that trained on genuinized data. However, if the attacker applying any geniunization, using GMM trained on the genuinized spoof data leads to almost perfect (and sometimes even perfect) classification. It emphasis the need of a detector whether the data genuinized or not. Other observation about these experiments are: for no action case, using GMMs on genuinized data harm more using CQCC features for LA conditions, and LFCC features for PA conditions; for other cases, CQCC features perform always better than LFCC features. The results withe both GMMs trained on randomly genuinized data (experiments $13-15$) do not match all the other experiment. We do not have good explanation for this phenomena. Only LA conditions with CQCC features behaved as accepted.

\begin{table}[!ht]
  \centering
   \caption{\label{table:GMM_GenuineGenuinized}{LA and PA EER [\%]: Repeating experiments $1-3$ from table \ref{table:LA_PA_GenuineGenuinized} with different GMMs.}}
  \begin{tabular}{r|cccc|cc||cc}
    & & & & & \multicolumn{2}{c}{\textbf{\textit{LA}}} & \multicolumn{2}{c}{\textbf{\textit{PA}}}\\
  	\hline
  	& $\bf{H\_Tr}$ & $\bf{S\_Tr}$ & $\bf{H\_Te}$ & $\bf{S\_Te}$ & \textbf{LFCC} & \textbf{CQCC} & \textbf{LFCC} & \textbf{CQCC}\\
	\hline
	$\bf{1}$ & O & O & NN & NN & $2.710$ & $0.430$ & $11.960$ & $9.870$\\
	$\bf{2}$ & O & O & NN & GN & $1.570$ & $3.336$ & $18.887$ & $11.924$\\
	$\bf{3}$ & O & O & NN & RN & $1.720$ & $4.749$ & $20.797$ & $11.630$\\
	\hline
	$\bf{4}$ & O & G & NN & NN & $34.380$ & $43.480$ & $30.220$ & $19.540$\\
	$\bf{5}$ & O & G & NN & GN & $0.080$ & $0.004$ & $2.074$ & $0.057$\\
	$\bf{6}$ & O & G & NN & RN & $0.042$ & $0.000$ & $1.425$ & $0.057$\\
	\hline
	$\bf{7}$ & G & G & NN & NN & $34.080$ & $43.210$ & $25.580$ & $19.91$\\
	$\bf{8}$ & G & G & NN & GN & $0.073$ & $0.002$ & $2.164$ & $0.133$\\
	$\bf{9}$ & G & G & NN & RN & $0.040$ & $0.000$ & $1.407$ & $0.037$\\
	\hline
	$\bf{10}$ & O & R & NN & NN & $37.050$ & $46.510$ & $31.030$ & $27.930$\\
	$\bf{11}$ & O & R & NN & GN & $0.082$ & $0.009$ & $1.277$ & $0.146$\\
	$\bf{12}$ & O & R & NN & RN & $0.044$ & $0.000$ & $0.813$ & $0.057$\\
	\hline
	$\bf{13}$ & R & R & NN & NN & $9.607$ & $35.675$ & $9.000$ & $16.070$\\
	$\bf{14}$ & R & R & NN & GN & $2.001$ & $0.080$ & $29.928$ & $34.407$\\
	$\bf{15}$ & R & R & NN & RN & $1.727$ & $0.110$ & $29.294$ & $35.222$\\
	\hline
  \end{tabular}
\end{table}

%% file: conclusions.tex
\section{Conclusion}
\label{sec:conclusions}

In this work, we presented a comparative analysis of time-domain related information of authentic and artificially modified speech in the context of speaker identification spoofing systems and associated countermeasures. In doing so, we aim to fill part of the gap that exists in the study of time domain information for the speaker identification spoofing domain. To cope with the inherent complexity of time domain approaches, we used a straightforward approach, which is perhaps the simplest possible: examine the \textit{probability mass function} (PMF) of the waveform coefficients. Our work was carried out within the framework of the ASVspoof 2019~\cite{ASVspoof2019} challenge and uses the datasets and core systems officially published for this occasion.

The first important result of this article is the confirmation of our initial hypothesis regarding the importance of time domain information for the speaker spoofing domain, thanks to the significant difference that we observed between the PMF of genuine speech and the PMF of the spoofed speech, especially at low amplitudes. At these amplitudes, the non-speech portions of the audio signal contribute significantly to the PMF waveform. This observation may explain the counter-logical but common practice of not using VAD, and therefore of using non-speech parts, in countermeasure systems. This finding prompts us to suggest paying more attention to the low-energy parts of the audio signal when doing text-to-speech conversion or dedicated speaker spoofing conversion, such as gaps between words or unvoiced segments.

In a previous work, we proposed a \textit{genuinization} algorithm in order to convert the samples of the spoofed speech, so that their new distribution becomes similar to the genuine speech PMF.
The genuinization algorithm is based on continuous distributions, algorithm \ref{alg:Genuanization}, and works well for LA conditions. Applying this algorithm to data for PA conditions did not yield good results and revealed a gap in the PMF, as shown in Figure \ref{fig:trainDataPMFs_PA}. The reason seems to be due to very high probabilities around the zero value of the PA conditions PMF  (illustrated in figure \ref{fig:Spoof2GenuineConversion_21}). To solve this problem, a $5$-bit perturbation was added (illustrated in figure \ref{fig:Spoof2GenuineConversion_31}). The new algorithm works well for PA conditions, figure \ref{fig:trainDataPMFs_PA_Fixed}, and LA conditions.

After a qualitative inspection of the time domain PMF and successful genuinization of the spoofed speech, the next step was to test the effect of genuinization on the countermeasure system as provide by the ASVspoof 2019 challenge organizers. It was shown that both, the attacker and the countermeasure have to choose the strategy. For the attacker is to decide to genuinize the spoofed speech or not; while the countermeasure have to decide two issues: $1^{st}$, how to train the GMMs, with or without genuinization of the data; $2^{nd}$, decide whether to genuinized the received speech or not. From table \ref{table:LA_PA_GenuineGenuinized} it can be seen that it is better that the countermeasure will not genuinize the received speech regardless from the decision of the attacker, when the GMMs where trained on the original data. This fact stayed valid in all other cases as well. Another interesting observation presented in experiments $4-9$. Although the performances are not good, it is important to see that if the countermeasure action is to genuinized the received speech, it practically, regardless the attacker action the EER almost does not affected. This observation is also valid for all the other cases of other GMMs training possibilities.

In table \ref{table:GMM_GenuineGenuinized} the presented results were for different GMMs training without taking any action on the received speech. It is clear that if the attacker do not genuinize the spoofed speech, the best countermeasure strategy is to stay with the original models (experiment $1$). If the attacker choose to implement either genuinization or random genuinization, original models are far to be the best choice. Excluding the case of LA conditions with LFCC features, the increase in the EER is significant (experiments $2$ and $3$). However, even for LA conditions with LFCC features, the results are far to be the best that can be achieved. The best solution is to train the GMM of the spoof speech on genuinized or randomly genuinized data. The genine speech GMM important not to train with randomly genuinized data (experiments $13-15$) while all the other options are practically equivalent (experiments $4-12$). In all these experiments the EER for genuinized speech is very low, especially for the CQCC features, while the EER is very high in case the attacker do not genuinized the speech (experiments $4$, $7$, and $10$). It emphasizes the necessity of an anti-spoofing system that can generalize the both cases.

To conclude, much work should be done in several directions: $1^{st}$, the spoofed data must take into account not only the spectral information, but also the waveform information; $2^{nd}$, the fact that non-speech is so important for countermeasure and and also, so different in the waveform, points that it is poorly presented in the spoofed speech and have to be better described; $3^{rd}$, even the very simple time domain manipulation can dramatically harm the countermeasure performance and it important to developed systems that are immune to such manipulations; $4^{th}$, we presented a very simple time domain manipulation, but much work still to be done, such as taking better care for non-speech event and taking time dependencies into account and not only $1^{st}$ order statistic.

To conclude, it is only the first work in this direction and there is much to do by combining the time and frequency domains to a common for both, spoofing and anti-spoofing.